\documentstyle[prb,eqsecnum,aps,epsf]{revtex}
\begin{document}
\draft
\setlength{\textfloatsep}{10pt plus 2 pt minus 2 pt}
\twocolumn[\hsize\textwidth\columnwidth\hsize\csname@twocolumnfalse%
\endcsname
\title{The density matrix renormalization group method.\\
Application to the PPP model of a cyclic polyene chain.}
\author{G. Fano, F. Ortolani$^{\dag}$ and L. Ziosi}
\address{
  Dipartimento di Fisica, Universit\'a di Bologna\\
  Via Irnerio 46 , 40126 Bologna -  Italy\\
  fano@bologna.infn.it\\
  $^{\dag}$Present address: International Centre for Theoretical Physics,\\
  Trieste, Italy}
\date{\today}
\maketitle

\begin{abstract}

The density matrix renormalization group (DMRG) method  introduced by White
for the study of strongly interacting electron systems is reviewed;
the method is variational and considers a system of localized electrons
as the union of two adjacent fragments $A$,$B$. A density matrix $\rho$
is introduced, whose eigenvectors corresponding to the 
largest eigenvalues  are  the most
significant, the most probable states of $A$ in the presence
of $B$; these states are retained, while  states corresponding 
to small eigenvalues
of $\rho$ are neglected.    It is conjectured that the
decreasing behaviour of the eigenvalues is gaussian.  The DMRG method is
tested on the Pariser-Parr-Pople Hamiltonian of a cyclic polyene
$ (CH)_N$ up to $N=34$. A Hilbert space of dimension $5. \times 10^{18}$
is explored. The  ground state energy is  $ 10^{-3} eV $
within the full CI value
in the case $N=18$. The DMRG  method compares favourably also with
coupled cluster approximations.
The unrestricted
Hartree-Fock solution (which presents spin density waves)
is briefly reviewed, and a comparison is made with the DMRG energy values.
Finally, the spin-spin and density-density correlation functions are
computed; the results suggest that the antiferromagnetic order of the
exact solution does not extend up to large distances but exists locally.
No charge density waves are present.

\bigskip
\end{abstract}

]
\narrowtext

\section{Introduction}

     A few years ago White \ \cite{white1} introduced in the study of electron
correlation a new and powerful numerical method: the density
matrix renormalization group (DMRG). The method provided
extremely accurate results in the case of the one-dimensional
Heisenberg and Hubbard models \ \cite{white1,white2,white3,qin,noack},
Hubbard-like models with bond alternation\ \cite{pang-liang}
and recently has been applied to some two dimensional models
\ \cite{xiang,white2d}.\par
       
     DMRG is a new variational method that promises to be very 
useful in quantum chemistry.
It deals with the main difficulty of this kind of calculations,
i.e. the exponential increase of the dimension of the Hilbert
space with the size of the  system, in a new, direct and
efficient way. While the usual packages of ab initio quantum
chemistry cut the dimension of the Hilbert space by neglecting
the coefficients of the configuration interaction expansion
below a certain threshold, the DMRG obtains an analogous result with
a different strategy.
A system of localized electrons is partitioned
in two blocks $A$,$B$  (sometimes $B$ is called "environment"
or "universe") and all the many-electron states corresponding
to situations in which the population of the two blocks is unphysical
(e.g. all electrons in $A$, no electrons in $B$) are automatically
truncated by the formalism.
A density matrix is introduced, whose eigenvectors,
corresponding to the larger eigenvalues, are the
most significant, the most probable states of $A$ in the presence
of $B$.

These  states are retained, and states  corresponding to very
small eigenvalues are neglected. The two blocks are taken
initially small and increase their size in the course of the
calculation. As a result of the systematic truncation mentioned
above, the time of computation 
does not grow more than the fourth power 
of the size of the system, keeping constant the number $m$ of
retained eigenvectors of the density matrix.
The error \cite{liang-pang} is an exponentially decreasing function of $m$.
\par

      The method is especially suited to treat systems with
translational or reflection invariance, since in an intermediate
stage of the calculation wave functions suitable to describe
the block $B$  can be obtained simply 
by translation (or reflection) from those of  block $A$.\par

      A good candidate in order to test the method in quantum chemistry
is provided by the Pariser-Parr-Pople model of conjugated polyenes.
Many considerations are in favour of this choice:

\begin{itemize}

\item A cyclic polyene $ (CH)_N$ with the carbon atoms at the vertices
of a regular polygon is "translationally" invariant (here translation
means a rotation of the circle circumscribed to the polygon); hence
the simplification mentioned above can be applied.\par

\item Exact full configuration interaction calculations are available
\  \cite{stef1,stef2},
 and we can compare the DMRG ground state energy
values with these results. The comparison can be made up to $N=18$. A
further comparison can be made with the coupled cluster
(CC) method \ \cite{paldus1}. However, the DMRG method is much more powerful;
we have computed without much effort ground state energy values
up to  $N=34$ carbon atoms. The full CI Hilbert space corresponding
to $N=34$ has dimension equal to ${34\choose 17}^2\approx 5.44\times 10^{18}$.
\par

\item Trans-polyacetylene presents interesting experimental and
theoretical problems: the bond alternation (and in particular
the values of the two bond lengths) can in principle be deduced
by ab initio computations but this problem meets considerable
difficulties. 
Recently an interesting approach to the problem of the
dimerization of polyacetylene using the DMRG method has been put
forward by 
M.B.Lepetit and G.M.Pastor \ \cite{lepetit};  these last authors
treat accurately the hopping term allowing a dependence on the
distance between the  $C$ atoms and describe the
electron interaction by a Hubbard term.
Therefore it would be of interest to extend their work
by substituting a PPP interaction to the Hubbard interaction.
In the present paper we show that this extension is possible
(but we do not derive the hopping term from ab initio
calculations). 
\par

\item The unrestricted Hartree-Fock solutions of the PPP model
Hamiltonian present spin density waves and charge density
waves \ \cite{paldus2,paldus3,paldus4,fukutome}.
It is of interest to know whether or not
these waves persist after a more precise variational approximation
to the ground state (like the DMRG) is performed.\par

\end{itemize}

      The paper is organized as follows: in Sec.2  the PPP
Hamiltonian is written down and the
DMRG method is reviewed. In particular we
point out  some mathematical aspects of the DMRG method that usually
are not sufficiently emphasized.  In Sec.3 the properties
of the  unrestricted  (spin density wave)
Hartree-Fock solution are briefly discussed. In Sec.4 the numerical
results and the conclusions are presented.

\section{The PPP Hamiltonian and the DMRG method.}

        The Pariser-Parr-Pople Hamiltonian of the $\pi$ electronic model
of a cyclic polyene $C_N H_N$ can be written as \ \cite{paldus1,paldus2,parr}:
\begin{equation}
        H = \beta \sum_{ < \mu \nu >} \hat{E}_{\mu \nu}
          + {1\over 2} \sum_{\mu,\nu=0}^{N-1} \gamma_{\mu \nu}
           \left(\hat {n}_\mu - 1 \right)
           \left( \hat {n}_\nu - 1 \right)
\end{equation}
where $\hat{E}_{\mu \nu}$  are the generators of the unitary
group summed over spin, and $\hat {n}_\mu = \hat {E}_{\mu \mu}$
is the occupation number of the site $\mu$; $\beta$,
$\gamma_{\mu \nu}$ are parameters of the model, and
$<\mu \nu >$ denotes summation restricted to nearest neighbor.
We limit ourselves to the series $N=2n=4 \nu +2, \nu=1,2,...$,
where $N$ denotes the total number of electrons which is equal
to the total number of sites.
According to ref. [\onlinecite{paldus1}] we take  $\beta = -2.5$ eV, and for the
Coulomb repulsion we use the Mataga-Nishimoto prescription \ \cite{mataga}:
\begin{equation}
        \gamma_{\mu \nu} = {1 \over  \gamma_0^{-1} + d_{\mu \nu}}
        \qquad\hbox{(a. u.)}
\end{equation}
where  $\gamma_0 = 10.84$ eV, $d_{\mu \nu}$ denotes the distance
between the vertex $\mu$ and the vertex $\nu$ of a regular polygon
of $N$ sites and is given by
\begin{equation}
        d_{\mu \nu} = b\,{ \sin(\mu-\nu){ \pi\over N}\over\sin{\pi \over N}}
\end{equation}
and $b$, the nearest-neighbor separation, is equal to $ 1.4$  \AA.\par
\smallskip

      Let's see now how the DMRG method can be applied to the
PPP model. We will also review briefly the principal formal
and physical ideas\ \cite{white1,white2,delgado}
 that are behind this new and powerful
numerical method.\par

      Let $A$ and $B$ denote two adjacent subsets of respectively
$N_A$, $N_B$ sites. The method consists of two parts: step 1, called
the "infinite system method" and step 2, called the "finite system
method".  In step 1, $N_A + N_B < N$, $N_A = N_B$  and  $N_A$, $N_B$
are progressively increased up to reach the condition $N_A + N_B = N$,
while in step 2, we have always $N_A + N_B = N$, with variable $N_A$ and
$N_B$.   For instance in step 1 we can have $N=18$,
$A=\{1,2,..6\}$, $B=\{7,8,..12\}$, in step 2 we can have
$A=\{1,2,3,4\}$, $B=\{5,6,....18\}$.
The main task of the method is to find a reduced set of ``localized''
many particle states for subsets (blocks)
 $A$ and $B$ suitable to describe
the union $A \bigcup B$.
\par

      Let us denote by $A^+$, $B^+$  polynomials in the creation
operators corresponding to sites in $A$, $B$, respectively.     Let
$|0\rangle$ denote the vacuum, and let
$|a\rangle = A^+|0\rangle,\ |b\rangle = B^+|0\rangle$.
Clearly $|a\rangle$, $|b\rangle$ represents states of electrons localized in
different subsets. We can form the state  $ A^+ B^+|0\rangle$;
this state is similar but not identical to the tensor product
$ |a\rangle \otimes \  |b\rangle $ since the operators $A^+$, $B^+$ do not
necessarily commute. We use the notation $ |a\rangle |b\rangle $
to denote the compound state  $A^+  B^+  |0\rangle$.
Clearly, varying the polinomials $A^+$, $B^+$ in all possible independent
ways, the states $|a\rangle|b\rangle$ generate the whole Hilbert space.
\par

\begin{table}
\caption{Energy results: the energies (in eV) calculated 
via Restricted  and Unrestricted HF, FCI\protect\tablenotemark[1] 
and DMRG are compared for different values of N. $m_A^{(n)}$ 
indicates the number of states kept in block A during the 
the n-th DMRG iteration.}
\begin{tabular}{ccccccc}
N&$E_{RHF}$&$E_{UHF}$&$E_{FCI}$\tablenotemark[1]
&$E_{DMRG}$&$m_A^{(1,2)} $&$m_A^{(3)}$ \\
\tableline
6&-11.358325&-11.358325&-12.722033&-12.722032&256&512\\
10&-17.441467&-17.910422&-20.060503&-20.060503&256&512\\
14&-23.731302&-24.924267&-27.671391&-27.671333&256&512\\
18&-30.101389&-32.007998&-35.385430&-35.384861&256&512\\
22&-36.513220&-39.105943&-&-43.145027 & 256 & 512\\
26&-42.950070&-46.207715&-&-50.928028 & 256 & 512\\
30&-49.403281&-53.310920&-&-58.715323 & 200 & 400\\
34&-55.867856&-60.414852&-&-66.509902 & 200 & 400\\
\end{tabular}
\tablenotetext[1]{From Ref. [\onlinecite{stef1,stef2}].}
\label{table1}
\end{table}

      Suppose that we have found an exact or approximate
ground state $|\psi\rangle$ of $ N_A + N_B $ electrons in the
subset $A \bigcup B$ of the chain;
let us expand  $ | \psi\rangle $ as:

\begin{equation}
        |\psi\rangle = \sum_{ I J} \psi_{ I J}  A^+_{I} B^+_{J} |0\rangle
\end{equation}
\noindent where $\{ A^+_{I} |0\rangle \} $ denote a complete orthonormal 
set of states
of electrons localized in $A$, and
$ \{ B^+_{J} |0\rangle \}$ 
is an analogous complete set of states corresponding
to $B$. For instance, initially we can have
$ A^+_{I} B^+_{J} |0\rangle = a^+_{i_1} a^+_{i_2} ... a^+_{N_A}
b^+_{j_1} b^+_{j_2} ... b^+_{N_B} |0\rangle $ where the $a^+$ , $b^+$
create electrons in $A$, $B$ respectively ; in this case the numbers
$\psi_{I J} $  are the usual configuration interaction (CI)
coefficients.
In principle the sums  $ \sum_{ I} $, $ \sum_{ J} $ run
over $ 4^{N_A} $, $ 4^{N_B} $ states respectively, since the occupation
numbers  $ n_\uparrow, n_\downarrow $ of a site can have four possible values:
$ (0,0), (1,0), (0,1), (1,1) $.  However the number of spin up electrons
and the number of spin down electrons are good quantum numbers and can be
fixed; we can choose states $A_I^+|0\rangle$, $B_J^+|0\rangle$ with fixed
numbers of spin up and spin down electrons, and
the coefficients   $ \psi_{ I J} $ vanish unless this
conservation law is fulfilled. Furthermore, during the iteration
procedure, the number of states will be truncated; therefore in the
expansion (2.4) we keep in general only $m_A$ states for the
block $A$ and $m_B$ states for the block $B$.
 In the following we shall assume
that the coefficients $\psi_{ I J}$ are real.\par

        The main mathematical tool of the DMRG theory is provided
by the following density matrix:
\begin{equation}
        \rho_{ I I'} = \sum^{m_B}_J  \psi_{I J} \psi_{ I' J}
                 = \left( \psi  \psi^T \right)_{ I I'}
\end{equation}
         The dimension of the matrix $ \rho$ is
$ m_A \times m_A$; however, because of the number
conservation laws described above, the matrix is actually in
block form: the number of up and down electrons of the
states $I$ and $I'$  must be the same. Furthermore  $\rho$
is a non negative square matrix. \par

         Let us first make some simplifying assumptions, that will
be relaxed in the following. Let's assume that the blocks
$A$ and $B$ are described by the same number of states ($m_A$ = $m_B$), so
that the matrix $\psi$ is a square matrix. Denoting by $S$ the
square root of $\rho$ ($\rho = \psi \psi^T = S^2$, $S = \rho^{1\over 2}$),
we have the polar decomposition
\begin{equation}
        \psi = S U_1
\end{equation}
where $U_1$ is an orthogonal matrix.  We diagonalize $S$ by writing
$ S = U D U^T $, where $U$ is an orthogonal matrix and $D$
is diagonal. Therefore we can write
\begin{equation}
        \psi = U D U^T U_1  = U D V^T
\end{equation}
where $V$ is an orthogonal matrix, and $ \rho = U D^2 U^T $.\par

         Actually formula (2.7) holds for {\it any} rectangular matrix
$m_A \times m_B$  $\psi$ (see e.g [\onlinecite{book}]). $U$ and $D$ are square
matrices $m_A \times m_A$ and  $V^T$ is $m_A \times m_B$. These matrices
verify the conditions:
\begin{eqnarray}
        U^T U = I\;;  \quad V V^T = I \;; \cr\cr
              \psi \psi^T = U D^2 U^T \;; \quad
              \psi^T \psi = V D^2 V^T
\end{eqnarray}
and $D$ is diagonal and non-negative. Let us denote by $D_\alpha$ the
eigenvalues of $D$.  Substituting (2.7) into (2.4) we obtain:
\begin{equation}
        | \psi\rangle =
       \sum^{m_A}_\alpha D_\alpha |u_\alpha\rangle |v_\alpha\rangle
\end{equation}
where
\begin{equation}
        |u_\alpha\rangle  = \sum_I^{m_A} U_{I \alpha} A^+_{I} 
        |0\rangle \;, \quad
        |v_\alpha\rangle =  \sum_J^{m_B} V_{J \alpha} B^+_{J} |0\rangle \;
\end{equation}
        What is the meaning of $ |u_\alpha\rangle , |v_\alpha\rangle$?
They  represent  states of the subsystems $A$, $B$, such that the
probability for the whole system  $A \bigcup B$ to be found in the
state $|u_\alpha\rangle  |v_\alpha\rangle$  is $ D_\alpha^2 $. The main idea of
the DMRG method consists in neglecting, in Eq.(2.9), all eigenvalues
$ D_\alpha $ below a certain threshold  which amounts to keeping
only a small number $m$ of terms in the sum (2.9) and using the
corresponding states $|u_\alpha\rangle$ as a basis for the description of
block $A$. Since
$ \hbox{Tr} \psi \psi^T  = \sum^{m_A}_\alpha D_\alpha^2 = 1$,
this approximation is good if the probabilities $ D_\alpha^2 $ have a
sufficiently rapid decrease to zero, so that  $ \sum_{\alpha=1}^m
D_\alpha^2 \simeq 1$.  At the best of our knowledge, all
numerical experiments performed so far (see, e.g. ref. 
[\onlinecite{lepetit,noi}])
confirm this rapid decrease of the probabilities $D_\alpha^2 $.
Let's give an heuristic argument for this decreasing behaviour.
Suppose that  $ A_I^+ $ creates $ N_A^e$ electrons in the $N_A$
sites of the block $A$, and $ B_J^+ $ creates $N_B^e$ electrons
in the $N_B$ sites of the block $B$.  In absence of the interaction,
usual statistical mechanics arguments prove that the probabilities
$D_\alpha^2 $ are strongly peaked about the populations  $N_A^e=N_A$,
$N_B^e = N_B$ (which correspond to a density of one electron
          per site); this is analogous to the classical result in statistical
mechanics stating that the probability of distributing
a large number of molecules in two communicating volumes is strongly
peaked about a distribution with equal density in the two volumes.
       
\begin{table}
\caption{Correlation energies: 
The correlation energies per electron (in eV)
of the FCI \protect\tablenotemark[1] and DMRG solutions
with respect to the Restricted  and Unrestricted  HF
approximations
are compared for different values of N.}\smallskip
\begin{tabular}[tb]{ccccccc}
  &\multicolumn{2}{c}{${{E-E(RHF)}\over N}$}&
   \multicolumn{2}{c}{${{E-E(UHF)}\over N}$}&\multicolumn{2}{c}{}\\ 
N & FCI\tablenotemark[1]& DMRG & FCI\tablenotemark[1] &DMRG 
&$m_A^{(1,2)} $&$m_A^{(3)}$ \\
\tableline
6&-0.227285&-0.227285&-0.227285&-0.227285&256&512\\
10&-0.261904&-0.261904&-0.215008&-0.215008&256&512\\
14&-0.281435&-0.281431&-0.196223&-0.196219&256&512\\
18&-0.293558&-0.293526&-0.187635&-0.187603&256&512\\
22&-&-0.301446&-&-0.183595&256&512\\
26&-&-0.306844&-&-0.181551&256&512\\
30&-&-0.310401&-&-0.180147&200&400\\
34&-&-0.313001&-&-0.179266&200&400\\
\end{tabular}
\tablenotetext[1]{From Ref. [\onlinecite{stef1,stef2}].}
\label{table2}
\end{table}

Because of the central limit theorem, the peak is gaussian in the
classical  case; we make the conjecture that even in the quantum
interacting case that we are considering, this gaussian behaviour
still holds, at least for translationally invariant systems, like
the PPP model.  Of course if the conjecture is true, it explains
the strongly decreasing behaviour of the probabilities $D_\alpha^2$
mentioned above.\par

      Let's now proceed with the description of the DMRG method.
Once we have a pretty good basis of $m_A$ states
$|u_\alpha\rangle$ that describe the block $A$,
and $m_B$ states $|v_\alpha\rangle$ that describe the block $B$,
the next task consists in the enlargement of the blocks.
In the first part of the algorithm (infinite system method),
since $N_A = N_B$ and the system is translationally invariant,
the states  $ |v_\alpha\rangle$  can be simply obtained by
translating the states $ |u_\alpha\rangle $.  Hence we can
concentrate our attention on the block $A$.\par

      The simplest way of enlarging the block $A$ consists in
adding a site $s$ to $A$, obtaining a new block  $ A' = A \bigcup s$.
White \ \cite{white1} denotes by $A \;\bf{\bullet}$ this new block. Denoting
by $ |s_1\rangle = |0\rangle$, 
$ |s_2\rangle = a_{s \uparrow}^\dagger |0\rangle$,
$|s_3\rangle = a_{s \downarrow}^\dagger |0\rangle$,
$|s_4\rangle = a_{s \uparrow}^\dagger   a_{s \downarrow}^\dagger |0\rangle$,
the states describing the site $s$, we have
$ 4 m_A $ vectors

\def\In{\lower2pt\hbox{$\scriptstyle I$}}

\def\Inn{\lower2pt\hbox{$\scriptscriptstyle I$}}
\def\Innp{\lower2pt\hbox{$\scriptscriptstyle {I'}$}}

\begin{eqnarray} 
	A'^+_I\,|0\rangle &= |u_{\alpha_{\Inn}}\rangle 
	|s_{\gamma_{\Inn}}\rangle,\qquad  I=(\alpha_{\In},\gamma_{\In})\cr\cr
	\alpha_{\In}&= 1,...m_A,\qquad\gamma_{\In} = 
	1,...4
\end{eqnarray}
in order to describe  $A' = A \;\bf{\bullet}\;=\; A \bigcup s$.\par
                                                                     
       At the same time, we add an analogous site  $t$  to the block $B$,
and we consider the vectors $| v_{\beta}\rangle  |t_{\delta}\rangle$
($ \beta = 1,2,...m_B, \;\delta=1,2,3,4 $) in order to describe
the block  $ B' = B \;\bf{\bullet}\;=\; B \bigcup t$. With such a basis
we can now proceed to compute the expansion (2.4) for the wavefunction
for the new superblock $A' \bigcup B'$.\par

       Let us use the term "local" to denote operators
$ a_{\mu}^\dagger , a_{\mu}, n_{\mu}$ referring to one site $\mu$ only,
and the term "internal to  block $A$" to denote operators whose site
indices belong to the  block $A$. \par
         The idea is now to compute a new "effective" Hamiltonian
matrix $H'$, by using the truncated basis consisting of the
$ 16 m_{A} m_{B} $ vectors  $ | u_{\alpha} \rangle | s_{\gamma} \rangle
| v_{\beta} \rangle | t_{\delta} \rangle $.

\begin{table}
\caption{Comparison of approximate solutions:
The correlation energy ${E-E_{RHF}}\over N$ (in eV)
of the DMRG solution is compared with
the partial cluster analysis $(|e^D\rangle|RHF\rangle)$ 
\protect\tablenotemark[1],
the Approximate Coupled Pair theory with Quadruples (ACPQ)\protect
\tablenotemark[2]
and the Approximate Coupled Pair theory with Triples and 
Quadruples (ACPTQ)\protect\tablenotemark[2].}
\begin{tabular}{ccccccc}
N & $|e^D \rangle|RHF\rangle\tablenotemark[1]$ & ACPQ
\tablenotemark[2] &
ACPTQ\tablenotemark[2] 
& DMRG &$m_A^{(1,2)}$&$m_A^{(3)}$\\
\tableline
 6& -0.224196& -0.2238  & -0.2253  &-0.227285 &256&512\\
10& -0.248723& -0.2515  & -0.2577  &-0.261904 &256&512\\
14& -0.256777& -0.2649  & -0.2762  &-0.281431 &256&512\\
18& -        & -0.2720  & -0.2887  &-0.293526 &256&512\\
22& -        & -0.2763  & -0.2994  &-0.301446 &256&512\\
\end{tabular}
\tablenotetext[1]{From Ref. [\onlinecite{stef2}].}
\tablenotetext[2]{From Ref. [\onlinecite{paldus1}].}
\label{table3}
\end{table}

Clearly it is easy to compute
terms of the Hamiltonian containing local operators referring
only to one of the four blocks  $ A,\;s,\;B,\;t $; these terms are known
from previous steps of the iteration. A little more
care is needed in order to compute terms like  $ a_{\mu}^\dagger a_{\nu}$
or  $ n_{\mu} n_{\nu}$ with $\mu,\;\nu$ belonging to different
blocks (e.g. $ \mu \in A, \;\nu \in s $, etc.). For this purpose
it is necessary to keep in the computer memory all the matrix
elements of the local operators  $ \langle u_{\alpha_{1}} |
a_{\mu}^\dagger | u_{\alpha_{2}} \rangle $,  
$ \langle v_{\beta_{1}} |n_{\nu} | v_{\beta_{2}} \rangle $.\par

        The entire procedure can now be repeated: we look for
the ground state vector $\psi'$ of the truncated Hamiltonian $H'$,
by using Lanczos's or Davidson's algorithm.  A new density matrix
$ \psi' {\psi'}^T $ and new state vectors $ |u_{\alpha}' \rangle $,
that represent states of $A'$, are computed according to
the analogous of the first of formulas (2.10)
which now reads:
\begin{eqnarray}
	|u'_\alpha\rangle &=& \sum_{I=1}^{4m_A} U'_{I\alpha} 
			A'^+_I|0\rangle \cr\cr
                     &=& \sum_{I=1}^{4m_A} U'_{I\alpha}
                                |u_{{\alpha}_{\Inn}}\rangle
				|s_{\gamma_{\Inn}}\rangle 
                                \quad\alpha = 1,...m_{A'}
\end{eqnarray}
Again we do not keep all the vectors: $m_{A'} $ is generally less than
$ 4m_A$ and  often one
puts  $ m_{A'} = m_A $, although this choice is not necessary.
The corresponding
$ |v_{\beta}' \rangle $ that describe $ B' $ are obtained from the
$ |u_{\alpha}' \rangle $ by translation.  \par
                                                  
In this new truncated basis we compute the matrix
elements of all the local and internal operators 
relative to block $A'$
and we keep them in the
computer memory, in order to use them in next steps of the method.
If, for example, we have an operator $O$ internal to block $A$,
it is also internal to the new block $A'$ and we have the following rule
to update its matrix elements:
\begin{eqnarray}
  \langle u'_{\alpha_1} | O | u'_{\alpha_2} \rangle  =
        \sum_{I,I' = 1}^{4m_A}
         U'_{I \alpha_1 } \langle 0 | A'_I O  A'^+_{I'}|0\rangle
                         U'_{I' \alpha_2 }\cr
	=\sum_{I,I'= 1}^{4m_A}
              U'_{I \alpha_1 } 
        \langle u_{\alpha_{\Inn}} | O | u_{\alpha_{\Innp}}\rangle
        \langle s_{\gamma_{\Inn}} | s_{\gamma_{\Innp}}\rangle
               U'_{I' \alpha_2 }\cr
        =
        \sum_{I,I'= 1}^{4m_A}
              U'_{I \alpha_1} 
        \langle u_{\alpha_{\Inn}} | O | u_{\alpha_{\Innp}}\rangle
        \delta_{\gamma_{\Inn} \gamma_{\Innp}}
               U'_{I' \alpha_2},\cr\cr
	\hbox{for}\quad\alpha_1,\alpha_2 = 1,...m_{A'}  
\end{eqnarray}
\par

Two more sites are added to the blocks
$A'$, $B'$, giving rise to new blocks  $ A'' = A' \ \bf{\bullet} $,
$ B'' = B' \ \bf{\bullet} $, etc.   By the systematic procedure
of adding two more sites, truncating the basis and updating
the Hamiltonian matrix at each iteration, systems of large
size can be handled. \par
                           
       A comment is in order about the choice of the two
sites that are added and their position with respect to blocks $A$ and $B$.
We can form the superblock
$A\;{\bf\bullet}\;B\;{\bf\bullet}$
or the superblock
$A\;{\bf\bullet}\;{\bf\bullet}\;B$.
White suggests that the enlarged configuration
$A\;{\bf\bullet}\;B\;{\bf\bullet}$
is to be preferred to
$A\;{\bf\bullet}\;{\bf\bullet}\;B$
in the case of periodic boundary conditions,
the opposite holds in the case
of open boundary condition. 
$ A \ {\bf{\bullet}}  B \ {\bf{\bullet}} $,
\vskip-30pt
\begin{figure}
\vskip-30pt
\centerline{\epsfxsize = 8.6cm\epsfbox{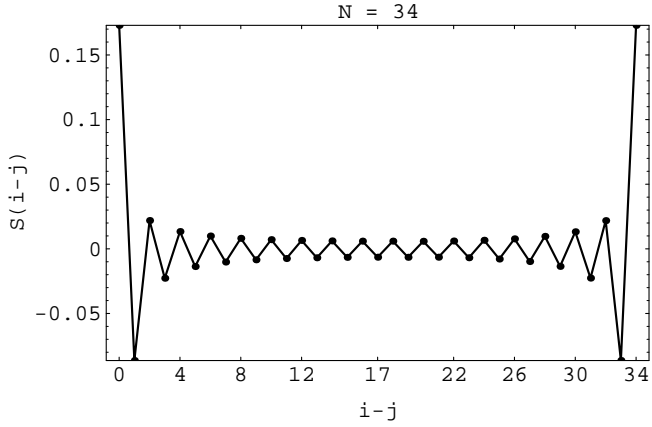}}
\vskip-30pt
\caption{Spin-spin correlation function versus
the distance between sites.\label{fig:s_di_ij}}
\end{figure}
\bigskip
 In fact the blocks $A$
and $B$ are separated by the site $t$ in the case
while they become adjacent by periodicity in
$ A \ { \bf{\bullet \  \bullet} } \  B $.
The kinetic part of the Hamiltonian (2.1) ``connects'' two blocks only
by its border sites with operators $a_\mu^\dagger$, $a_\nu$, whose
matrix elements are known.
 These matrices are ``big'' for blocks $A$ and $B$,
and ``little'' for the 1-site blocks $s$ and $t$, so the matrix elements of the
hamiltonian $H'$ are simpler when a ``big'' block is surrounded by 1-site
blocks. \par
              
      The "infinite system algorithm" is stopped when the number of sites
of  $A \bigcup B $ reaches the total number $N$ of sites.  In order
to improve the accuracy of the method, White himself proposed
a second algorithm, that we will briefly describe. This second
algorithm takes place after the infinite system algorithm
reaches the end. \par

        In the finite system algorithm, to an increase of $A$ by
one site it corresponds a {\it decrease} of the "universe"
$B$ by one site.
Denoting by $A_x$, $B_y$ blocks $A$ and $B$ with $x,y$ sites
respectively, we start with the system
$A_{ {N\over 2} - 1}\;{\bf\bullet}\;B_{ {N\over 2} - 1}\;{\bf\bullet}$
and we want to construct the systems
$A_{ {N\over 2}}\;{\bf\bullet}\;B_{ {N\over 2} - 2}\;{\bf\bullet}$,
$A_{ {N\over 2}+1}\;{\bf\bullet}\;B_{ {N\over 2} - 3}\;{\bf\bullet}$,
etc.
Therefore in order to use the translational invariance,
we need to keep in the computer memory all the relevant matrix elements
of   $A_{ {N\over 2} - 2 }$, $ A_{ {N\over 2} - 3 } $, etc.,
in order to be able to use the symmetry and produce the matrix
elements of  $B_{ {N\over 2} - 2 } $, $ B_{ {N\over 2} - 3 } $, etc.
It should be noticed that when $ m_{B'} < m_{A'} $  (this certainly
happens when $B$ becomes small) the rows of $\psi'$ cannot be
linearly independent. As a consequence, $\psi'{\psi'}^T$ has many eigenvalues
equal to zero. 

From Eq.(2.8) we see that the $ m_{A'} \times m_{A'} $
matrix  $ \psi' {\psi'}^T $ and the smaller $ m_{B'} \times m_{B'} $
matrix  $ {\psi'}^T \psi' $ have the same non vanishing eigenvalues.
In practice it is sufficient to diagonalize only the smallest of the two
density matrices.
The procedure stops when we reach the system
$ A_{N-3}  { \bf{\bullet \ }} B_1 {\bf{\bullet} }  $, i.e. when the block
$ B $
has reduced to a single site.
We can now increase $B$ and decrease $A$; the
subsystems $A$, $B$ behave like if they were separated by a
moving zipper.
At every step we increase the accuracy of the states
$|u_\alpha\rangle$ that describe the blocks $A_x$ and
after few oscillations of the zipper all the
blocks $A'_x, \  2 \le x \le N-2 $, accurately represent parts of
a complete system of $N$  sites, the remaining "universe" being
the corresponding $B'_{N-x}$ block.

During this procedure,
all the relevant matrix elements of the local operators
must be stored and updated.
A more detailed explanation of this point can be found in the
original paper by White \ \cite{white2}.
Usually one stops when $A$ and $B$ have the same length.\par

\section{ The unrestricted Hartree-Fock (UHF) solution.}

Let us still denote
by $ a_{\mu \sigma}^+ $ the creation operator of an electron
of spin $\sigma$ on the site $\mu$. The creation operator of an electron
in a symmetric Bloch orbital is given by   (we use 
letters $k, k_1, k_2..$ to denote the symmetric  orbitals):
\begin{equation}   
	a_{k \sigma}^+  =  {1\over \sqrt{N}} \sum_{k=0}^{N-1}
                   e^{i \omega k \mu}  a_{\mu \sigma}^+
                 \qquad  k=0,1,... N-1   
\end{equation} 

where $ \omega = {2 \pi \over N }$.  In terms of these operators,
the Hamiltonian can be written as
\begin{eqnarray} 
	H = 2 \beta \sum_{k \sigma} \cos(\omega k) a_{k \sigma}^+
	a_{k \sigma}  - E_0\cr\cr 
	+ {1\over 2} \sum_{k_1 k_3 k \sigma \tau}  K(k)
  	a_{k1 \sigma}^+  a_{k_3+k, \tau}^+  a_{k_3, \tau} a_{k_1+k,\sigma}
 \end{eqnarray}
where all $k$ indices run from $0$ to $N-1$, the constant term
$ E_0 = \sum_{\gamma < \nu} \gamma_{\mu \nu} $ has been added to
the Hamiltonian and represents the internuclear repulsion energy,
and $ K(k) $ is given by:
\begin{equation}   
	K(k)  =       {1\over {N}} \sum_{\mu}^{N-1}  \gamma_{0 \mu}
                   e^{i \omega k \mu}  
                 \qquad   k=0,1,... N-1   
 \end{equation} 
       Due to the discrete rotational symmetry of the polygon,
all indices can be taken modulo $N$. It is convenient 
to represent the $k$ indices on a circle  (see fig. 1 of
 ref. [\onlinecite{paldus4}]).
The restricted Hartree-Fock orbitals are determined by the
condition:
\begin{equation}  
	N-\nu  <  k  <  N+\nu   
\end{equation}
which characterizes the Fermi sea  $F$.  The restricted Hartree-Fock
(RHF)single particle energies are given by\cite{paldus3}:
\begin{equation} 
	\epsilon_{k} = 2 \beta \cos(\omega k) + N K(0)
	- \sum_{k_{1} \in F}  K(k - k_{1})  
  \end{equation}
and the total RHF energy is:
\begin{equation} 
	E_{RHF}  = -E_0   + \sum_{k\in F} [ 2 \beta \cos(\omega k) 
   + \epsilon_k ]  
 \end{equation}
      It is known since long time that it is possible to lower
the RHF ground state energy by considering molecular orbitals
that are linear combinations of the orbitals $\phi_k$ and
$\phi_{k+n} $ corresponding to two endpoints of a diameter
of the circle of Fig.1 in ref. [\onlinecite{paldus4}].\par

      Furthermore, taking into account also the spin indices of
the two orbitals, there are many different possibilities
that give rise to local minima of the UHF energy.  All these
possibilities have been carefully studied many years ago
by Fukutome \cite{fukutome}, Paldus and Cizek \cite{paldus3},
 and others, and
give rise to charge density waves (CDW) and spin density
waves (SDW).  However in the case of the Mataga-Nishimoto
prescription for the two center Coulomb repulsion integral
and with the values of the parameter given in Sec.1, we
have checked that the lowest UHF energy is obtained by the
following  BCS-Bogoliubov canonical transformation
(which corresponds to the $(A^{t} + B^{t})^+ $ case
of ref. [\onlinecite{paldus3}]):
\begin{eqnarray} 
    \gamma_{k \uparrow} &=  \ u_{k}  a_{k \uparrow}
                            + v_{k}  a_{k+n \uparrow }  \cr\cr 
      \gamma_{k \downarrow}&= - u_{k}  a_{k \downarrow}
                            + v_{k}  a_{k+n \downarrow } 
 \end{eqnarray}
where $ u_{k}^2 + v_{k}^2 = 1 $ , $ u_{k+n} = u_{k} $,
$ v_{k+n} = - v_k $ . The operators $\gamma_{k \sigma}^+ $
create UHF  orbitals, since the linear combination depend
on $\sigma$.\par

       The first-order density matrix (in the 
pseudomomenta representation) is given by:
\begin{equation}  
\langle a_{k_1 \sigma}^+  a_{k_2 \sigma} \rangle =
     \delta_{k_1,k_2}  f^{(1)}(k_1)  +
     \delta_{k_1,k_{2}+n} \ f^{(2)}(k_1) 
         (-1)^{\sigma} 
 \end{equation}
where  $ \ f^{(1)}(k) = u_k^2 $,  
       $ \ f^{(2)}(k) = u_k v_k $  for $k \in F$, and
       $ \ f^{(1)}(k) = v_k^2 $,  
       $ \ f^{(2)}(k) = - u_k v_k $  for $k \not\in F$.  In the 
original atomic-orbital basis we have the interesting formula:
\begin{equation}  
\langle a_{\mu \sigma}^+ a_{\mu \sigma} \rangle =
             {1 \over 2}  + (-1)^{\mu + \sigma}  \delta 
 \end{equation}
where $ \delta =  {1\over N} \sum_{k=0}^{N-1} |u_k v_k| $.  This 
formula shows the existence of SDW of antiferromagnetic type; the
occupation numbers $\langle n_\uparrow \rangle $ and  $ \langle n_\downarrow
 \rangle $
are different from each other on the same site (when one of the
two is larger than $ {1\over 2}$ the other is smaller than $ {1\over 2}$)
 giving rise to a decrease
of the on-site Coulomb repulsion.  Furthermore no CDW
appear, since   $ \langle n_\uparrow \rangle  + \langle n_\downarrow \rangle
 = 1 $.
\par

      The expectation value of the Hamiltonian (2.5) can be easily
computed by using Wick's theorem; minimization of $\langle H\rangle$ with 
respect to the coefficients  $u_k, v_k$ gives rise to the
following well known set of equations\cite{paldus3,fukutome} 
of the BCS type:
\begin{equation} 
	u_k^2 = {1\over 2} \left( 1 + { |\xi_k| \over \sqrt{ \xi_k^2
	+ \Delta_k^2 } } \right),\quad
   	v_k^2 = {1\over 2} \left( 1 - { |\xi_k| \over \sqrt{ \xi_k^2
	+ \Delta_k^2 } }\right) 
 \end{equation}
where  $ \xi_k = {1\over 2} ( \hat{\epsilon} (k) -
\hat{\epsilon} (k+n) )$, and the UHF orbital energies are given
by
\begin{eqnarray} 
	\hat{\epsilon} (k) = 2 \beta \cos(\omega k) + N K(0)
         - \sum_{q\in F} K(k-q) u_q^2\cr\cr
      - \sum_{q\not\in F} K(k-q) v_q^2, 
 	\qquad\hbox{for} \  k=0,1,...N-1
 \end{eqnarray}
and $\Delta(k)$ must fulfil the famous "gap equation":
\begin{eqnarray}  
	\Delta(k) =& {1\over 2} \sum_{q\in F}
      [ K(k-q) + K(k-q+n) ] \cr\cr
	&\times { \Delta(q) \over
      \sqrt{ \xi_q^2 + \Delta(q)^2 } },\qquad	
	\hbox{for} \ k=0,1,...N-1
 \end{eqnarray}
      If the only solution  of the gap equation is the trivial
solution $\Delta(k)=0$, we obtain simply the RHF ground state.
If a non trivial solution exists, the non linear system of equations
(3.10),(3.11),(3.12) can be easily solved numerically by
an iterative method. Starting with $ \Delta(q)=\hbox{constant} $
and $ \hat{\epsilon}(k) = \epsilon(k) $, we solve the
gap equation (3.12) by iteration. Usually $30-40$ iterations will
suffice. The solution  $\Delta(k)$ is substituted into
Eqs.~(3.10), (3.11); in this way we obtain a set of approximate
orbital energies $\hat{\epsilon}_1(k)$. The entire procedure
is repeated substituting  in the right hand side of Eq. (3.12)
the solution  $\Delta(k)$ and $\xi_k = 
{1\over 2} (\hat{\epsilon}_{1}(k) - \hat{\epsilon}_{1} (k+n)) $, etc.,
until the entire set of equations is fulfilled with sufficient
accuracy.\par

      The UHF ground state energy of the model is given by
\begin{eqnarray}
 E_{UHF} = 
  - \sum_{k_{1},k_{2}=0}^{N-1}\, K(k_2 -k_1)\cr\cr
           \times  \Big[ f^{(1)}(k_1) f^{(1)}(k_2)  
           +   f^{(2)}(k_1) f^{(2)}(k_2)\Big]\cr\cr
 -E_0 + {1\over 2}\, K(0)\, N^2  
 + 4 \beta \sum_{k \in F}\, \cos(\omega k) \,f^{(1)}(k)\cr\cr
 \end{eqnarray} 
       
\smallskip

The antiferromagnetic long-range order of the UHF solution
appears also in the height of the peak of the magnetic structure
factor:
\begin{eqnarray}   
	 &S(k) = \sum_{j=0}^{N-1}  e^{-i { 2 \pi \over N } j k }
              \langle  S_{z} (j) S_{z} (0) \rangle \cr\cr
     &= {1\over 4} + N \delta_{k,{N \over 2}}  \delta^2 
      - {1\over {2N}} \sum_{q=0}^{N-1} f^{(1)} (q) f^{(1)} (q-k)\cr\cr&
      - {1\over {2N}} \sum_{q=0}^{N-1} f^{(2)} (q-k-n) f^{(2)} (q) 
  \end{eqnarray}
which is reached for $k={N\over 2} $.  We have :
\begin{equation}   
	S( {N \over 2}) = {1\over 4} +
         { 1 \over N }  \big[ \sum_q  | u_q v_q | \big]^2
        -{ 1 \over N }    \sum_q (u_q v_q )^2      
 \end{equation}
and this quantity scales like $N$ for large $N$.

\begin{figure}
\centerline{\epsfxsize=8.6cm\epsfbox{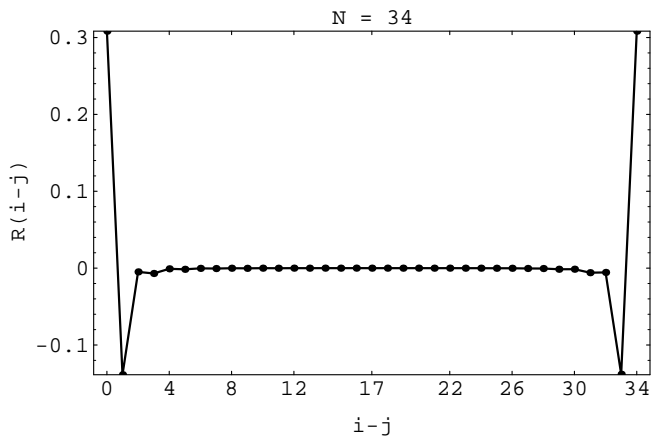}}
\vskip-30pt
\caption{ Density-density correlation function
versus  the distance between sites.\label{fig:r_di_ij}}
\end{figure}\noindent

\section{Numerical results and conclusions }
In Table \ref{table1} we  show the energy  
results calculated with the DMRG method 
up to $N = 34$, and we compare them 
with RHF, UHF and FCI energies
(the FCI energies are available only up to $N=18$).\par
We see that the relative error of the DMRG solution
with respect to FCI 
is only $2.1 \times 10^{-6}$ for $N=14$ and $1.6 \times 10^{-5}$  
for $N=18$, which is a
quite satisfactory  result.\par
Table \ref{table2} shows 
the correlation energy per electron of the FCI and DMRG solutions
with respect to the RHF and UHF approximations. \par
       The DMRG method compares favourably with the Coupled
Cluster method;   in Table \ref{table3} the correlation energies
  $(E - E_{RHF}) / N$    are compared with coupled cluster results of
ref. [\onlinecite{stef2,paldus1}].
  The DMRG energy is slightly lower  than  the
Approximate Coupled Pair with Triples and Quadruples (ACPTQ) value.
    All calculations were performed  
iterating the DMRG algorithm three times 
(the first iteration uses the
infinite
system method, the second and third iterations use the finite system method).
We  stop when A and B 
have the same length.\par
In the first iteration  the size
of the system grows, but the potential between two sites is
kept equal to its final value, i.e. to the value attained
 when the number of sites of the polygon is N.  
Generally we keep 256 states in 
block A during both the first and the second
iteration; in  
 order to achieve a better
convergence, during 
the third iteration we keep 512 states.
In the heaviest 
calculations ($N=30,34$) we keep
only $200-400$ states in block A, due to 
memory/disk-space limitations. It should be noted that the disk-memory
requirement grows with the number of sites even if the number of
retained states  
is held constant. This is  due to the long range  nature
of the interaction 
that forces us to keep on disk a linearly growing number of matrices
that represent the local operators.
We have checked that the disk-space grows as $N m^2$.\par
 In Fig. \ref{fig:s_di_ij}
the spin-spin correlation function
$S(i-j) = \langle S_z(i)S_z(j)\rangle$
is plotted : a short range antiferromagnetic
order is clearly present. We have computed the Fourier transform $S(k)$
which of course reaches its maximum value for $k={N\over 2}$, like the
UHF-SDW solution (see (3.14)). 
 However, the growth is linear with $N$ for the
UHF-SDW solution, but scales approximately as $ 0.1398 + 0.1457 \hbox{Log} N $
for the DMRG solution.  Therefore we cannot speak of long range
SDW.  Also the CDW are ruled out by the present calculation. This can
be seen from the graph of the density-density correlation function    
       \begin{equation}
            R(i,j) =  \langle n(i) n(j) \rangle - 
                          \langle n(i) \rangle
                          \langle n(j) \rangle 
       \end{equation}
(see Fig. \ref{fig:r_di_ij}). \par
       Concluding, the DMRG method provides a very powerful tool for
the calculation of energies and properties of simple many electrons
Hamiltonians. 
It gives results very close to full CI results
and is able to handle Hilbert spaces of very large dimension.
 It would be of great interest to apply the method
to a realistic many electrons Hamiltonian, possibly after a previous
localization of the occupied and virtual orbitals.  However, this
program meets with
some difficulty because of the large number of matrices that
must be kept when the four orbitals of the interaction term 
belong to different blocks.

\acknowledgments
 The authors are greatly indebted to
G.L. Bendazzoli for teaching them the peculiarities of the PPP model,
and to A. Parola for extremely useful discussions and suggestions.


\end{document}